\documentclass[aps,prb]{revtex4}%
\usepackage{amsfonts}
\usepackage{amsmath}
\usepackage{amssymb}
\usepackage{epsfig}
\usepackage{graphicx}

\begin{document}

\title{DNA Confined in Nanochannels: Hairpin Tightening by Entropic Depletion}
\author{Theo Odijk*, Complex Fluids Theory, Kluyver Laboratory and of
Biotechnology, Delft University of Technology, and Julianalaan 67,
2628 BC Delft, the Netherlands}

\begin{abstract}
A theory is presented of DNA hairpins enclosed in a nanochannel. A
hairpin becomes constrained as it approaches the wall of a channel
which leads to an entropic force causing the hairpin to tighten.
The free energy of the hairpin computed in the classical limit is
significantly larger than what one would expect. As a result, the
distance between hairpins or the global persistence length is
often tens of $\mu$m long and may even reach mm sizes for $10$ nm
thin channels. The hairpin shape and size, and the DNA elongation
are computed for nanoslits, and circular and square nanoschannels.
A comparison with experiment is given. \vspace{10pt}

*E-mail address: odijktcf@orange.nl
\end{abstract}

\maketitle

\section{Introduction}

There has been a longstanding interest in the effects of confining
polymer chains in tubes and slits, for instance, in relation to
the field of size exclusion chromatography
\cite{CAS1,CAS2,COL,DAO,CAN,GUI,ODI1,GOR}. Recent advances in chip
lithography have led to the fabrication of well-defined
nanochannels \cite{GUS, TEG1} so that issues in polymer physics
may now be addressed which were elusive in the past
\cite{BAK,TEG2,REI,CHO,KOS,MAN,FAN,BAL,RIE}. At the same time,
stiff polymers like DNA and actin can now be well-aligned which is
of considerable interest in the biosciences \cite{RAN,JO}.

In order to understand the mechanism of chain elongation, it is crucial to
investigate the alignment as a function of the size of the channel cross
section and this has been carried out recently by Reisner et al \cite{REI}.
The channel width was of the order of the DNA persistence length. They
focused especially on the relaxation times of monodisperse double-stranded
DNA fluctuating in square and rectangular nanochannels which they compared
with estimates inferred from theory \cite{DAO,ODI1}. The pictures of
fluorescently labeled DNA they show are, however, quite remarkable with
respect to their extreme degree of elongation, which has not been remarked
upon before. Let us consider, in fact, a naive estimate of the distance $g$
between two consecutive hairpins of the nanoconfined DNA which would be
something like%
\begin{equation}
g=\pi r\exp \left( U_{0}/k_{B}T\right)  \label{VGL1}
\end{equation}%
The radius of a hairpin, which is supposed to be semicircular, is $r$ and
its bending energy is $U_{0}$. The temperature is $T$\ and $k_{B}$\ is
Boltzmann's constant. A form similar to eq (\ref{VGL1})\ arises in the
theory of nematically confined stiff chains \cite%
{KHO,GEN,WAR,GLA,VRO,ODI2,SPA} and will be rederived below. Note
that the B-helix is assumed to remain intact. Thus, in the Reisner
experiment, in the case of DNA confined within a $60\times 80$ nm
channel for instance, one would expect the hairpin to extend to
its maximum diameter of 100 nm (equal to the diagonal). This leads
to $g\simeq 1$ $\mu$m from eq (\ref{VGL1}). This is the mere step
length of the effectively one-dimensional random walk, yet the T2
DNA having a contour length of $63$ $\mu$m is almost fully
elongated \cite{REI}!

I argue here that there is an important effect missing in eq (\ref{VGL1}),
namely entropic depletion of the DNA hairpin by the wall of the nanochannel.
As the hairpin approaches the wall, it loses orientational and translational
degrees of freedom. The resulting entropic repulsion forces the hairpin
toward the axis of the channel so the hairpin is tightened up. Eq (\ref{VGL1}%
) seriously underestimates the elongation for this reason. My emphasis will
be on an analysis of this phenomenon in the elastic or classical limit; an
analytical solution of the relevant Fokker Planck equation for a stiff chain
trapped in a pore is well-known to be difficult although numerical analyses
have been performed \cite{DIJ,BUR,BIC,CHE}. The hairpin is thus assumed to
be a two-dimensional curve within a plane aligned along the long axis of the
nanochannel. The hairpin chord, which is the vector distance between the two
ends, is perpendicular to this axis but the shape of the curve is not
semicircular and needs to be determined. The chord of length $2r$ has to be
fitted into the cross section of the channel. This is a geometrical problem
in two dimensions. Clearly, the confinement of the chord entails a loss in
both orientational and translational degrees of freedom, giving rise to a
free energy $F_{conf\text{ }}$of entropic origin. This decreases as the
hairpin tightens up so it balances the bending energy $U$ at an optimum
radius $\bar{r}$. Accordingly, the free energy of the hairpin in the
classical limit is%
\begin{equation}
\bar{F}_{cl}=\bar{U}\left( \bar{r}\right) +\bar{F}_{conf}\left( \bar{r}%
\right)  \label{VGL2}
\end{equation}%
where a bar denotes the state of minimum free energy.

My main objective will be to compute this quantity. It is convenient to
express the scale $g$ as%
\begin{equation}
g=\alpha \bar{r}\exp \left( F_{tot}/k_{B}T\right)  \label{VGL3}
\end{equation}

\begin{equation}
F_{tot}=\bar{F}_{cl}\left( \bar{r}\right) +H  \label{VGL4}
\end{equation}%
Here, $\alpha $ is a constant and $H$ is a sum of a variety of fluctuational
free energies beyond the classical limit which I will analyze qualitatively.

The paper is organized as follows. In the next section, I compute
the optimum shape of the hairpin and the minimum bending energy.
In section 3, the free energy of confinement is evaluated for
three types of cross section: circular, square and rectangular.
The distance\ between hairpins or the global persistence length is
calculated in section 4. In the last section, a discussion will be
given of the resulting free energies, hairpin sizes and DNA
elongations in comparison with experiment. Therein, the terms
neglected in the classical limit will also be remarked upon.

\section{Hairpin shape}

The hairpin has a length $l$ which is to be determined while its end-to-end
distance is constrained to be $2r$. It is described by the tangential unit
vector $\overrightarrow{u}\left( s\right) $ at contour position $s$ (see fig
1). The hairpin curve is assumed to be symmetric about the $y$ axis which is
parallel to the central axis of the nanochannel. The angle between $%
\overrightarrow{u}\left( s\right) $ and the x axis is $\pi -\omega \left(
s\right) $ so that the unit vector may be written in terms of the coordinate
unit vectors $\overrightarrow{u}=-\overrightarrow{e}_{x}\cos \omega +%
\overrightarrow{e}_{y}\sin \omega $. The boundary conditions are assumed to
be $\omega \left( 0\right) =\frac{1}{2}\pi $ and $\omega \left( l/2\right)
=0 $.

I first compute the shape of the curve by minimizing the bending
energy of the hairpin of length $l$
\begin{equation}
U=Pk_{B}T\int_{0}^{l/2}ds\left( \frac{d\omega }{ds}\right) ^{2}
\label{VGL5}
\end{equation}%
subject to the constraint%
\begin{equation}
\int_{0}^{l/2}ds\text{ }\vec{u}\cdot
\vec{e}_{x}=-\int_{0}^{l/2}ds\cos \omega =-r  \label{VGL6}
\end{equation}%
In eq (\ref{VGL5}), $P$ is the persistence length, $T$ is the temperature, $%
k_{B}$ is Boltzmann's constant and $d\omega /ds$ equals the inverse radius
of curvature. Minimizing eq (\ref{VGL5}) with respect to $\omega \left(
s\right) $ leads to%
\begin{equation}
\frac{d^{2}\omega }{ds^{2}}=\frac{\gamma }{2l^{2}}\sin \omega  \label{VGL7}
\end{equation}%
where the Lagrange multiplier $\gamma $ may be positive, zero or negative.
Yamakawa and Stockmayer \cite{YAM} have previously analyzed this equation in
their theory of the excluded-volume effect of stiff chains but our treatment
differs from theirs because the constraint is nonzero here (eq (\ref{VGL6}%
)). Integration of eq (\ref{VGL7}) yields%
\begin{equation}
l^{2}\left( \frac{dw}{ds}\right) ^{2}=f-\gamma \cos \omega  \label{VGL8}
\end{equation}%
where the constant $f$ must be positive.

In the remainder of the analysis we need not compute $\omega \left( s\right)
$ explicitly. The bending energy is as yet given in terms of three parameters%
\begin{equation}
\frac{U}{Pk_{B}T}=\frac{f}{2l}-\frac{\gamma r}{l^{2}}  \label{VGL9}
\end{equation}%
We ultimately wish to minimize this with respect to $l$ subject to eq (\ref%
{VGL6}). But it turns out that it is possible to do this analytically\ via
the dummy variable $B$ defined below.

Actually, there is another restriction on eq (\ref{VGL8}): the chain cannot
cross the wall of the channel. This is difficult to take into account
mathematically because it is a local constraint defined at every point of
the contour $s$. Nevertheless, on physical grounds, one expects deviations
of reverse curvature away from a typical hairpin configuration to be
forbidden in view of the increase in elastic energy associated with such
bending. Accordingly, I assume the simpler constraint $d\omega /ds<0$ holds.

The right hand side of eq (\ref{VGL8}) is now rewritten in terms of the new
angular variable $v=\left( \pi -\omega \right) /2$ with boundary values $%
v\left( 0\right) =\pi /4$ and $v\left( l/2\right) =\pi /2$.%
\begin{equation}
l\frac{d\omega }{ds}=-\left( f+\gamma \right) ^{\frac{1}{2}}\left( 1-B\sin
^{2}v\right) ^{\frac{1}{2}}  \label{VGL10}
\end{equation}%
The constant $B<1$ is defined by%
\begin{equation}
B\equiv \frac{2\gamma }{f+\gamma }  \label{VGL11}
\end{equation}%
Note that $B$ may be positive or negative though the combination $f+\gamma $
cannot be negative. Next, the objective is to the express the bending energy
given by eq (\ref{VGL9}) solely in terms of $B$. First, I integrate eq (\ref%
{VGL10}) over the entire contour from $0$ to $l/2$. This leads to%
\begin{equation}
\left( f+\gamma \right) ^{\frac{1}{2}}=4\xi  \label{VGL12}
\end{equation}%
in terms of the elliptic integral%
\begin{equation}
\xi \equiv \int_{\pi /4}^{\pi /2}\frac{dv}{\left( 1-B\sin ^{2}v\right) ^{%
\frac{1}{2}}}  \label{VGL13}
\end{equation}%
The constraint eq (\ref{VGL6}) is rewritten as%
\begin{equation}
r=-\int_{0}^{\pi /2}d\omega \left( \frac{ds}{d\omega }\right) \cos
\omega  \label{VGL14}
\end{equation}%
The derivative is now substituted from eq (\ref{VGL10}) and $\cos \omega $
is expressed in terms of $v$. The scaled contour length $\alpha \equiv l/r$
thus becomes a function explicitly dependent on $B$ only.%
\begin{equation}
\alpha ^{-1}B\xi =\xi \left( 1-\frac{1}{2}B\right) -\beta  \label{VGL15}
\end{equation}%
where $\beta $ is another elliptic integral
\begin{equation}
\beta \equiv \int_{\pi /4}^{\pi /2}dv\left( 1-B\sin ^{2}v\right) ^{\frac{1}{2%
}}  \label{VGL16}
\end{equation}%
The constants $f$ and $\gamma $ can also be written in terms of $B$ with the
help of eqs (\ref{VGL11}) and (\ref{VGL12}).%
\begin{equation}
f=8\xi ^{2}\left( 2-B\right)  \label{VGL17}
\end{equation}%
\begin{equation}
\gamma =8B\xi ^{2}  \label{VGL18}
\end{equation}%
We now need to derive an expression for the bending energy which may be
differentiated with respect to $B$ as expediently as is possible. Eqs (\ref%
{VGL17}) and (\ref{VGL18}) are first substituted into eq (\ref{VGL9}) and
then $B$ is eliminated with the help of eq (\ref{VGL15}).%
\begin{equation}
E\equiv \frac{Ur}{Pk_{B}T}=\frac{8\xi ^{2}}{\alpha }\left( 1-\frac{1}{2}%
B-B\alpha ^{-1}\right) =\frac{8\beta \xi }{\alpha }  \label{VGL19}
\end{equation}%
This is still too complicated because it contains $\alpha $. Hence, I
eliminate $\alpha $ by inserting eq (\ref{VGL10}) into eq (\ref{VGL14}) and
using eq (\ref{VGL12}). Eq (\ref{VGL19}) then reduces to the form%
\begin{equation}
E\left( B\right) =4\eta \beta  \label{VGL20}
\end{equation}%
in terms of the integral%
\begin{equation}
\eta \equiv \int\_{\pi /4}^{\pi /2}dv\frac{g\left( v\right)
}{\left( 1-B\sin ^{2}v\right) ^{\frac{1}{2}}}  \label{VGL21}
\end{equation}%
and the function%
\begin{equation}
g\left( v\right) \equiv 2\sin ^{2}v-1  \label{VGL22}
\end{equation}%
The advantage of writing the energy as the product $\eta \beta $ is that the
numerator and denominator are identical in the respective integrands so that
one may expect cancellation of terms after differentiation (apart from the
factor $g$ which, not inconveniently, increases monotonically with $v$). In
fact, the derivative of eq (\ref{VGL20}) with respect to $B$ introduces $%
\sin ^{2}v$ into the integrands which is then conveniently rewritten in
terms of $\left( 1-B\sin ^{2}v\right) $. Terms of the kind $\eta \beta $
then cancel and we obtain%
\begin{equation}
\frac{1}{2}\frac{dE}{dB}=B^{-1}\left[ \beta \int_{\pi /4}^{\pi /2}dv%
\frac{g\left( v\right) }{\left( 1-B\sin ^{2}v\right) ^{3/2}}-\xi \eta \right]
\label{VGL23}
\end{equation}%
This form is reexpressed as a two-dimensional integral with the help of eqs (%
\ref{VGL13}), (\ref{VGL16}) and (\ref{VGL21}). After rearrangement, this
yields%
\begin{equation}
\frac{1}{2}\frac{dE}{dB}=\int_{\pi /4}^{\pi /2}\int_{\pi
/4}^{\pi /2}dudv\frac{g\left( u\right) \left( \sin ^{2}u-\sin ^{2}v\right) }{%
\left( 1-B\sin ^{2}u\right) ^{3/2}\left( 1-B\sin ^{2}v\right) ^{\frac{1}{2}}}
\label{VGL24}
\end{equation}%
Next, one notes that the function $g\left( u\right) /\left( 1-B\sin
^{2}u\right) $ increases monotonically with $u$ for all $B$, whatever the
sign. The remaining factor in the integrand is antisymmetric in $u$ and $v$.
Hence, the derivative must be positive for all $B$.

In conclusion, the bending energy is a minimum as $B\rightarrow -\infty $.
Asymptotically, we have from eqs (\ref{VGL13}) and (\ref{VGL16})%
\begin{equation}
\xi \sim C_{1}\left( -B\right) ^{-\frac{1}{2}}  \label{VGL25}
\end{equation}%
with%
\[
C_{1}=-\ln \tan \frac{\pi }{8}=0.881374
\]%
\begin{equation}
\beta \sim C_{2}\left( -B\right) ^{\frac{1}{2}}  \label{VGL26}
\end{equation}%
with%
\[
C_{2}=\frac{1}{2}\sqrt{2}
\]%
Eq (\ref{VGL15}) yields for the optimum length of the hairpin%
\begin{equation}
l/r=\alpha =\left( \frac{C_{2}}{C_{1}}-\frac{1}{2}\right) ^{-1}=3.3082
\label{VGL27}
\end{equation}%
The minimum elastic energy of the hairpin according to eq (\ref{VGL19}) is%
\begin{equation}
\frac{U_{m}r}{Pk_{B}T}=E_{m}=4\left( 1-C_{1}C_{2}\right) =1.5071
\label{VGL28}
\end{equation}

I have integrated eqs (\ref{VGL15}) and (\ref{VGL19}) numerically in order
to assess the nature of the elastic "well" (see table I). The optimum curve
is very close to that of a semicircle $\left( B=0;\alpha =\pi \right) .$The
bending energy only starts to increase markedly as $B$ approaches unity. The
hairpin is almost a straight line in that case, with highly rounded ends.

\section{Entropy of depletion}

In the Introduction it was argued that it is plausible to introduce a
classical limit for the state of the hairpin which is aligned along the
centerline of the nanochannel. Then, we need to address two two-dimensional
problems, the shape of the hairpin treated above and how the chord of the
hairpin can be fitted into the cross section of the channel. In the latter
case, we may speak of an entropy of depletion since orientational and
translational degrees of the chain are cut off as the hairpin is formed and
restricted by the channel walls. In this section I give approximate
expressions for the depletion entropy for three characteristic types of
nanochannels: circular, square and retangular.

\subsection{Circular cross section}

The chord of the hairpin is a segment of length $2r$ enclosed within a
circle of radius $a$ (see fig 2). The translational degrees of freedom of
the segment are assessed by how we may position point $P$ in the circle
which is at a distance $t$ from the origin $O$. The orientational degrees of
freedom are expressed by the angle $\theta $. In this subsection, all
lengths will be scaled by $a$ for convenience. Note that we require $t\leq
2r-1$ otherwise the segment does not fit into the circle. In addition, if $%
t>1-2r$, the segment cannot rotate freely and the maximum angle $\theta _{m}$
is given by%
\begin{equation}
1=4r^{2}+t^{2}-4rt\cos \theta _{m}  \label{VGL29}
\end{equation}%
It is in this regime that an approximation to the entropy is derived. The
fraction of realizable states is written as%
\begin{equation}
I_{c}=\left( 2\pi ^{2}\right) ^{-1}4\pi \int_{2r-1}^{1}dt\text{ }%
t\int_{0}^{\theta _{m}}d\theta  \label{VGL30}
\end{equation}%
This is simply a sum over all degrees of freedom noting that $\theta
_{m}=\theta _{m}\left( t\right) $. It is normalized by a factor $2\pi ^{2}$
applicable to the state if the segment were to orient and translate freely
within the circle.

The bending forces of the previous section tend to open up the hairpin
toward the channel wall. Therefore, let us assume $r$ and $t$ are close to
unity and introduce the small quantities $\delta $ and $\varepsilon $ : $%
\delta =1-r$ and $\varepsilon =1-t$ $\left( 0<\varepsilon <2\delta \right) $%
. In that case, the angle $\theta _{m}$ is also small and we may write $\cos
\theta _{m}\simeq 1-\frac{1}{2}\theta _{m}^{2}$. From eq (29) we have to the
leading order%
\begin{equation}
\theta _{m}^{2}=2\delta -\epsilon  \label{VGL31}
\end{equation}%
Performing the integrations in eq (30), we get the entropy of depletion%
\begin{equation}
S_{c}=k_{B}\ln I_{c}=\frac{3}{2}k_{B}\ln \text{ }\delta +k_{B}\ln \left(
\frac{8.2^{\frac{1}{2}}}{3\pi }\right)  \label{VGL32}
\end{equation}%
Although this expression is formally valid in the limit $\delta \ll 1$, it
turns out to be a very good approximation for all $r$, even when the segment
is virtually free to rotate. As $r=1-\delta \rightarrow 0$, $I_{c}\simeq
1.200$ is close to unity which is the exact value. Furthermore, $I_{c}$
divided by the integral computed numerically is a monotone increasing
function of $\delta $. The leading approximation to the integrand $t\theta
_{m}(t)$ in eq\ (\ref{VGL30}), namely $a\theta _{m}$ with $\theta _{m}$
given by eq (\ref{VGL31}), is also very near the numerical value of $t\theta
_{m}(t)$ for all $r>\left( 1-t\right) /2$.

\subsection{Square cross section}

Trying to fit a line segment into a square is different from the case
discussed above. If the side of the square has a length $A$ and the length
of the segment is restricted to the regime $A<2r<A\sqrt{2}$, then the
segment cannot rotate through 360 degrees (see fig 3). One of its ends
(denoted by point $P$) is constrained to lie inside the region that is
almost like a triangle in fig 3. Its boundary in terms of the Cartesian
coordinates $x_{b}$ and $y_{b}$ is given by%
\begin{equation}
x_{b}^{2}+y_{b}^{2}=4r^{2}  \label{VGL33}
\end{equation}%
Again, I attempt to approximate the entropy of depletion by a suitable
limiting form in this case, when the segment is almost as long as the
diagonal of the square. For convenience, all lengths are scaled by $A$ in
the rest of this subsection. Thus the small quantities $\delta _{s}$, $%
\varepsilon _{x}$ and $\varepsilon _{y}$ are introduced: $\delta _{s}=1-r%
\sqrt{2}$, $\varepsilon _{x}=1-x$ and $\varepsilon _{y}=1-y$ in terms of the
coordinates $x$ and $y$ of point $P$. To the leading order, we then have
from eq (33) the requirement%
\begin{equation}
\varepsilon _{x}+\varepsilon _{y}<2\delta _{s}  \label{VGL34}
\end{equation}%
The orientational degree of freedom is expressed by the angle $\theta $
which is bounded by the angles $\theta _{1}$ and $\theta _{2}$ (see fig 3).
They are given by%
\begin{equation}
\cos \theta _{1}=\frac{y}{2r}  \label{VGL35}
\end{equation}%
\begin{equation}
\sin \theta _{2}=\frac{x}{2r}  \label{VGL36}
\end{equation}

Since the segment approaches a diagonal in size, it is convenient to
introduce the small angles $\varphi _{1}$and $\varphi _{2}$ defined by $%
\varphi _{1}=\frac{1}{4}\pi -\theta _{1}$ and $\varphi _{2}=\theta _{2}-%
\frac{1}{4}\pi $ respectively. The left and right hand sides of eqs (35) and
(36) are now expanded in all of the small variables which yields $\varphi
_{1}=\delta _{s}-\varepsilon _{x}$ and $\varphi _{2}=\delta _{s}-\varepsilon
_{y}$ to the leading order. Hence, the range of the orientational degree of
freedom is%
\begin{equation}
\theta _{2}-\theta _{1}=2\delta _{s}-\varepsilon _{x}-\varepsilon _{y}
\label{VGL37}
\end{equation}%
Accordingly, the fraction of realizable states is, again, a sum over all
degrees of freedom%
\begin{equation}
I_{s}=\frac{2}{\pi }\int_{\text{triangle}}\int dxdy\left( \theta
_{2}-\theta _{1}\right) \simeq \frac{2}{\pi }\int_{0}^{2\delta
_{s}}d\varepsilon _{x}\int_{0}^{2\delta _{s}-\varepsilon
_{x}}d\varepsilon _{y}\left( 2\delta _{s}-\varepsilon
_{x}-\varepsilon _{y}\right) =\frac{8\delta _{s}^{3}}{3\pi }
\label{VGL38}
\end{equation}%
and the entropy of depletion is expressed as%
\begin{equation}
S_{s}/k_{B}=\ln I_{s}=3\ln \delta _{s}+\ln \left( \frac{8}{3\pi }\right)
\label{VGL39}
\end{equation}%
The number of corners is four which has to be included in the normalization
(because of fluctuations, the chord of the hairpin can, of course, sample
the entire square cross section of the nanochannel). It would appear that eq
(\ref{VGL38}) is a very good approximation to the exact $I_{s}$ because eq (%
\ref{VGL38}) equals 0.8488 at $\delta _{s}=1$ which is quite close to unity.
However, the numerically computed $I_{s}$ is more than twice that predicted
by eq (\ref{VGL38}) at $r=0.5$ so the approximation must be viewed as fairly
accurate at best for all $r<\frac{1}{2}\sqrt{2}A$.

\subsection{Slitlike cross section}

A simple approximation for the entropy of depletion can be given in the case
of a nanoslit of large aspect ratio i.e. when the cross section is a
rectangle of width $A$ very much larger than height $D$ (see fig 4a). We
suppose $D$ is so small that we may write $\theta _{m}\simeq x/r$ for the
maximum angle $\theta _{m}$ when the hairpin chord just touches one of the
walls (see fig 4b). Ultimately, this implies that the hairpin must be
sufficiently stiff ($\pi P\gtrsim D$, say). If the chord or line segment is
at a distance $x$ from the wall, the contribution to the orientational
entropy is proportional to $\ln \left( 2x/\pi r\right) $ owing to the
angular restriction ($\theta _{m}$ would be $\pi /2$ if the segment were to
orient freely). Hence, the total orientational entropy is given by%
\begin{equation}
S_{or}/k_{B}=\frac{2}{D}\int_{0}^{D/2}dx\ln \left( 2x/\pi r\right)
=\ln \left( \frac{D}{\pi r}\right) -1  \label{VGL40}
\end{equation}%
The segment may also translate along the longitudinal axis of length $A$
implying a fraction of realizable states equal to about $\left( A-2r\right)
/A$. The total entropy of depletion thus becomes%
\begin{equation}
S_{slit}/k_{B}=\ln \left[ \left( \frac{A-2r}{A}\right) \frac{D}{\pi r}\right]
-1  \label{VGL41}
\end{equation}%
This is only valid at small enough $D$. For this reason a smooth crossover
from eq (\ref{VGL41}) to eq (\ref{VGL39}) does not exist as $A\rightarrow D$
and $r\sqrt{2}\rightarrow A$.

\section{Global persistence length}

The distance between hairpins for a chain in a uniaxially ordered matrix
scales as the global persistence length which is given in terms of a
fluctuation theorem involving the chain susceptibility \cite{KHO,VRO,ODI2}.
This is a rigorous way of deriving $g$ but is only useful if one has a
precise analytical theory for the segmental distribution inside the
nanochannel, which is lacking at present. I have yet to establish eqs (\ref%
{VGL1}) and (\ref{VGL2}) and in the classical limit the following physical
argument is well known \cite{GEN,VRO}. The hairpins of length $\alpha \bar{r}
$ may be viewed as defects in a one-dimensional chain model of contour
length $L$. There are $L/g$ defects of energy $\bar{F}_{cl}=\bar{F}%
_{cl}\left( \bar{r}\right) $ and concentration $\alpha \bar{r}/g$.
Therefore, the free energy of the chain becomes%
\begin{equation}
\frac{F_{L}}{k_{B}T}=\frac{L}{g}\ln \left( \frac{\alpha \bar{r}}{g}\right) -%
\frac{L}{g}+\left( \frac{L}{g}\right) \frac{\bar{F}_{cl}}{k_{B}T}
\label{VGL42}
\end{equation}%
where the first two terms derive from the ideal entropy of the defect gas.
Minimization of the free energy with respect to $g$ leads to%
\begin{equation}
g_{cl}=\alpha \bar{r}\exp \left( \bar{F}_{cl}/k_{B}T\right)  \label{VGL43}
\end{equation}%
The free energy of confinement or depletion $F_{conf}=-TS$ has been
discussed in the previous section. If\ we set $S=0$, $U=U_{0}$ and we assume
the bend is\ semicircular, we regain eq (\ref{VGL1}). Any approximations
incurred in the previous analyses can be accounted for by adding an energy $%
H $ to $\bar{F}_{cl}$. The global persistence length will now be evaluated
for the various nanochannels.

\subsection{Circular nanochannel}

The bending energy of a hairpin is given by eqs (\ref{VGL19}) and (\ref%
{VGL28}), and eq (\ref{VGL32}) expresses the free energy of depletion for
the hairpin enclosed in the circular nanochannel. Upon minimization, the
total free energy of the hairpin in the classical limit%
\begin{equation}
\frac{F_{cl}^{c}\left( r\right) }{k_{B}T}=\frac{E_{m}P}{r}-\frac{3}{2}\ln
\left( \frac{a-r}{a}\right) -\ln \left( \frac{8.2^{\frac{1}{2}}}{3\pi }%
\right)  \label{VGL44}
\end{equation}%
leads to an optimum radius%
\begin{equation}
\bar{r}=\frac{1}{3}\left[ \left( E_{m}^{2}P^{2}+6E_{m}Pa\right) ^{\frac{1}{2}%
}-E_{m}P\right]  \label{VGL45}
\end{equation}%
Using eqs (\ref{VGL43})-(\ref{VGL45}), I have tabulated $\bar{r}$
and $g$ for DNA with a typical persistence length of $50$ nm (see
table II). These results will be discussed in the next section.

\subsection{Square nanochannel}

The entropy of depletion is now given by eq (\ref{VGL39}) so that the free
energy of a hairpin in a square nanochannel is%
\begin{equation}
\frac{F_{cl}^{s}\left( r\right) }{k_{B}T}=\frac{E_{m}P}{r}-3\ln \left( \frac{%
A-r\sqrt{2}}{A}\right) -\ln \left( \frac{8}{3\pi }\right)  \label{VGL46}
\end{equation}%
Minimization of $F_{cl}^{s}$ with respect to $r$ gives%
\begin{equation}
\bar{r}=\frac{1}{6}\left[ \left( E_{m}^{2}P^{2}+6\sqrt{2}E_{m}AP\right) ^{%
\frac{1}{2}}-E_{m}P\right]  \label{VGL47}
\end{equation}%
The optimum radius and distance between hairpins are displayed in
table III for DNA which has a persistence length of 57.5 nm. These
predictions will be compared with the experiments of Reisner et
al.

\subsection{Nanoslit}

Eq (\ref{VGL41}) leads to the free energy of a hairpin in a nanoslit%
\begin{equation}
\frac{F_{cl}^{slit}\left( r\right) }{k_{B}T}=\frac{E_{m}P}{r}-\ln \left[
\left( \frac{A-2r}{A}\right) \frac{D}{\pi r}\right] +1  \label{VGL 48}
\end{equation}%
Minimization yields an optimum radius%
\begin{equation}
\overline{r}=\frac{E_{m}PA}{A+2E_{m}P}  \label{VGL49}
\end{equation}%
Note that this expression does not depend on $D$. In this case, it is
possible to present a convenient formula for the distance between hairpins%
\begin{equation}
g=\frac{\pi \alpha E_{m}P\overline{r}}{D}\exp \left( \frac{2A+2E_{m}P}{A}%
\right)  \label{VGL50}
\end{equation}%
In a recent experiment \cite{JO}, the dimensions of the nanoslit
were kept fixed but the persistence length was varied by changing
the ionic strength. In table IV, $g$ is presented as a function of
$P$ for a $100\times 1000$ nm nanoslit.

\section{Discussion}

The first striking conclusion is that the distance between
hairpins is considerably larger than the usual persistence length
of DNA. This is true even when $P$ is smaller than the typical
width of the channel, irrespective of the type (see tables II, III
and IV). In the larger channels, the effect of entropic depletion
is clearly seen for the hairpins are tightened up significantly
(compare $r$ with $a$ or $A$). At the other extreme, the global
persistence length may reach the scale of a mm which means it may
prove feasible to align bacterial chromosomes in nanochannels
completely. Entropic depletion is an important factor in the
elongation.

The large values of $g$ are intimately related to the small size of the
orientational fluctuations of the aligned (nonhairpinlike) sections of the
chain with respect to the channel axis. This relation has been discussed at
length for a wormlike chain in a nematic \cite{ODI2}. Here, the mathematics
of a chain in a channel involves both translational and orientational
degrees of freedom so is significantly more complicated. Nevertheless, the
scaling estimate \cite{ODI1} for the angular fluctuations, when amended by a
small numerical coefficient (see eq (\ref{VGL53}) below), qualitatively
corroborates the large values of $g$ proposed here. A further implication is
that the crossover to the scaling regime applicable in the limit of flexible
chains \cite{DAO} is not trivial. The transition is delayed because $g$ is
still very large at $P\simeq $ channel width. The crossover must occur at $%
g\simeq P$ but, as I argue below, the classical limit is no longer valid
then.

It is interesting to note that the depletion effect is much stronger when
the nanochannel is square than when it is circular (compare the coefficients
1.5 and 3 in eqs (\ref{VGL44}) and (\ref{VGL46})). As a hairpin retracts
from the edge of a square channel, its freedom to move increases rapidly.
This is much less so in the case the channel is circular for the curvature
is directed smoothly inward then.

The entropic depletion by a nanoslit given by eq (\ref{VGL41}) is quite
peculiar. It leads to an expression for the optimum radius (eq (\ref{VGL49}%
)) which is independent of $D$. Moreover, for highly elongated slits, the
radius never increases beyond the persistence length. It would be
interesting to see how robust this classical computation is to fluctuations.
A priori one might expect a semiflexible chain confined in a very thin slit
of infinite dimensions to behave essentially as a two-dimensional chain of
persistence length $2P$, but this is not what is predicted here. The precise
range of validity of eq (\ref{VGL41}) needs to be established.

Reisner et al have presented fluorescent images of DNA in square (or almost
square) nanochannels \cite{REI}. The elongated length $R_{\exp }$ of the DNA
represents the typical average span of the chain and is shown in table V as
a function of the nanochannel size in the case of $T2$ DNA. The span of a
polymer chain is the smallest size of the box it can be enclosed in and in
one dimension its average has been computed in the flexible limit ($L\gg g$;
see refs \onlinecite{DAN} and \onlinecite{RUB})%
\begin{equation}
S=4\left( \frac{gL}{\pi }\right) ^{\frac{1}{2}}  \label{VGL51}
\end{equation}%
The average span is not known for the wormlike chain model but as $g$ tends
to $L$ or becomes larger, the root-mean-square extension $R_{e}$ should be
close to $S$.%
\begin{equation}
R_{e}^{2}=2Lg-2g^{2}\left( 1-e^{-L/g}\right)  \label{VGL52}
\end{equation}%
Besides the hairpins, orientational fluctuations also shorten the chain. If $%
\eta $ is the angle between its tangent and the axis of the channel, we have
the following average%
\begin{equation}
\langle \eta ^{2}\rangle =0.340\left( \frac{A}{P}\right) ^{2/3}
\label{VGL53}
\end{equation}%
This is derived from Burkhardt's expression \cite{BUR2}\ for the
distribution of a harmonically confined worm, suitably (though
approximately) rescaled to account for the hard wall repulsion in a square
nanochannel (see ref \onlinecite{JO}). In table V, the effective span ($%
S\left\langle \cos \eta \right\rangle $ or $R_{e}\left\langle \cos \eta
\right\rangle $) is compared with the experimental $R_{exp}$ \cite{REI}
where the cosine is computed to second order via eq (\ref{VGL53}). The
agreement is quite good if one bears in mind that no adjustable parameters
have been used. DNA of comparable size is also highly elongated in nanoslits
of dimensions presented in table IV. This essentially agrees with the large
values of $g$ computed from eq (\ref{VGL50}). We are currently investigating
whether this expression is quantitatively accurate \cite{JO}.

I finally discuss a variety of fluctuational terms that have been neglected
in the classical limit.

1) Eq (\ref{VGL44}) may be expanded to second order in order to evaluate
fluctuations in the chord or size of a hairpin. The root-mean-square
fluctuation is then $\langle \delta r^{2}\rangle ^{1/2}\simeq r^{2}/P$.

2) The shape of a hairpin fluctuates at constant $r$ . The length of the
hairpin is largest when the bending energy is a minimum. Away from the
minimum, the energy and length are shown in table I from which one estimates
the fluctuation in $l$ to be $\left\vert \delta l\right\vert \simeq r^{2}/P$.

3) The hairpin may tilt away from the channel axis but one expects the tilt
to be given by eq (\ref{VGL53}) or a similar expression when the cross
section is circular. Furthermore, the hairpin is only slightly more
eccentric than semicircular, at best, so the influence of tilting would
appear to be negligible.

4) The entropy of depletion has been computed only approximately when the
cross section of the nanochannel is square. In the case of DNA, this may
suffice because any correction terms to eq (\ref{VGL39}) will be sensitive
to the fluctuations in $r$ discussed above. However a complete analysis may
turn out to be useful for very stiff polymers like actin.

5) The hairpin curve is perturbed by thermal motion. The deformations are
undulatory in nature and may be expressed by $\Delta \simeq r^{3/2}/P^{\frac{%
1}{2}}$ \cite{ODI2,SHI,CLA}.

On the whole, we conclude that the classical limit breaks down as $r$
approaches $P$. For the widest channels in tables II and III, the
predictions for $g$ must be considered to be tentative only, for this
reason. Complete analyses of the global persistence length including
fluctuations have been given for chains in the nematic state \cite{VRO,SPA}.
It would be important to have similar rigorous treatments in the case of
nanochannels.

\subsection*{Acknowledgment}

I thank Peter Prinsen and David C. Schwartz for discussions.

\section*{Tables}

\

\begin{tabular}{|c|c|c|}
\hline
$B$ & $\alpha $ & $E$ \\ \hline
0.99999 & 2.0820 & 16.7402 \\ \hline
0.99 & 2.5101 & 2.7350 \\ \hline
0.8 & 2.9035 & 1.8418 \\ \hline
0.5 & 3.0491 & 1.6250 \\ \hline
0.1 & 3.1290 & 1.5772 \\ \hline
0 & 3.1416 & 1.5708 \\ \hline
-0,5 & 3.1846 & 1.5508 \\ \hline
-5 & 3.2705 & 1.5185 \\ \hline
-100 & 3.3058 & 1.5078 \\ \hline
-10000 & 3.3082 & 1.5071 \\ \hline
\end{tabular}

\bigskip

Table I. Scaled length and scaled energy E of the hairpin as a function of
B. At $B=0$ the hairpin is a semicircle so $\alpha =\pi $ and $E=\pi /2$.

\bigskip

\bigskip

\begin{tabular}{|c|c|c|c|c|c|c|c|c|c|}
\hline $a$ (nm) & 10 & 15 & 20 & 25 & 30 & 50 & 70 & 100 & 150
\\ \hline
$r$ (nm) & 8.6 & 12.1 & 15.3 & 18.3 & 21.1 & 31 & 39 & 50 & 65 \\
\hline $g$ ($\mu$m) & 2900 & 199 & 51 & 22 & 13 & 4.1 & 2.5 & 1.8
& 1.3 \\ \hline
\end{tabular}

\bigskip

Table II. Hairpin radius $r$ and global persistence length $g$ for
DNA confined in a circular nanochannel of radius $a$. The DNA
persistence length is 50 nm.

\bigskip

\bigskip

\begin{tabular}{|c|c|c|c|c|c|c|c|}
\hline
$A$ (nm) & 35 & 70 & 80 & 135 & 190 & 370 & 440 \\
\hline
$r$ (nm) & 15.9 & 26.0 & 28.5 & 40 & 50 & 74 & 82 \\
\hline $g$ ($\mu$m) & 317 & 27 & 19 & 7.0 & 4.4 & 2.5 & 2.3
\\ \hline
\end{tabular}

\bigskip

Table III. Hairpin radius $r$ and global persistence length $g$
for DNA confined in a square nanochannel with side equal to A. The
DNA persistence length is 57.5 nm. These results pertain to the
experiments of Reisner et al \cite{REI}.

\bigskip

\bigskip

\begin{tabular}{|c|c|}
\hline $P$ (nm) & ($\mu$m) \\ \hline 100 & 18
\\ \hline 150 & 42 \\ \hline 200 & 79 \\ \hline 250 & 32 \\ \hline
300 & 204 \\ \hline
\end{tabular}

\bigskip

Table IV. Global persistence $g$ of DNA enclosed in a nanoslit as
a function of the persistence length $P$. The width $A$ of the
nanoslit is 1000 nm and its height $D$ is 100 nm (values
pertaining to the experiments of Jo et al \cite{JO}).

\bigskip

\bigskip

\begin{tabular}{|l|c|c|c|c|c|c|c|}
\hline $A$ (nm) & 35 & 70 & 80 & 135 & 190 & 370 & 440 \\ \hline
$S$ ($\mu$m) &  &  &  & 47 & 38 & 28 & 27 \\ \hline $S\langle
\cos \theta \rangle$ ($\mu$m) &  &  &  & 33 & 23 & 12 & 9 \\
\hline $R_{e}$ ($\mu$m) & 61 & 46 & 41 &  &  &  &  \\ \hline
$R_{e}\langle \cos \theta \rangle$ ($\mu$m) & 54 & 37 & 33 &  & &
&  \\ \hline $R_{exp}$ ($\mu$m) & 47 & 42 & 35 & 23 & 13 & 10 & 8
\\ \hline
\end{tabular}

\bigskip

Table V. Experimental elongations $R_{exp}$ compared with
theoretical predictions $R_{e}\langle \cos \theta \rangle $ and
$S\langle \cos \theta \rangle $ in their respective regimes of
validity. $R_{exp}$ was measured directly from Fig 4a in ref
\onlinecite{REI}. The T2 DNA has a contour length of 63 $\mu$m.

\section*{Figures}

\vspace{40pt}

%The bounding box is usually not correct (see gsview and show bounding box).
%You have to adjust this manually in e.g. wordpad.

\begin{figure}[htbp]
  \begin{center}
      \epsfig{file=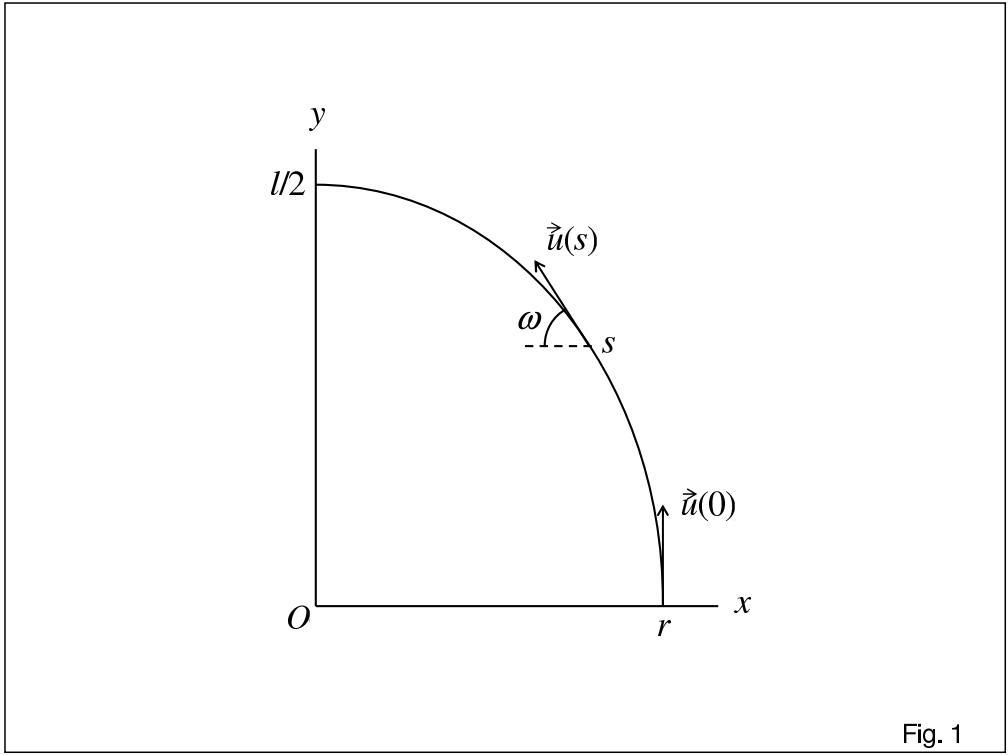, angle=270, width=380pt}
      \caption{Half of a hairpin curve in a two-dimensional Cartesian coordinate system.}
    \end{center}
\end{figure}

\begin{figure}[htbp]
  \begin{center}
      \epsfig{file=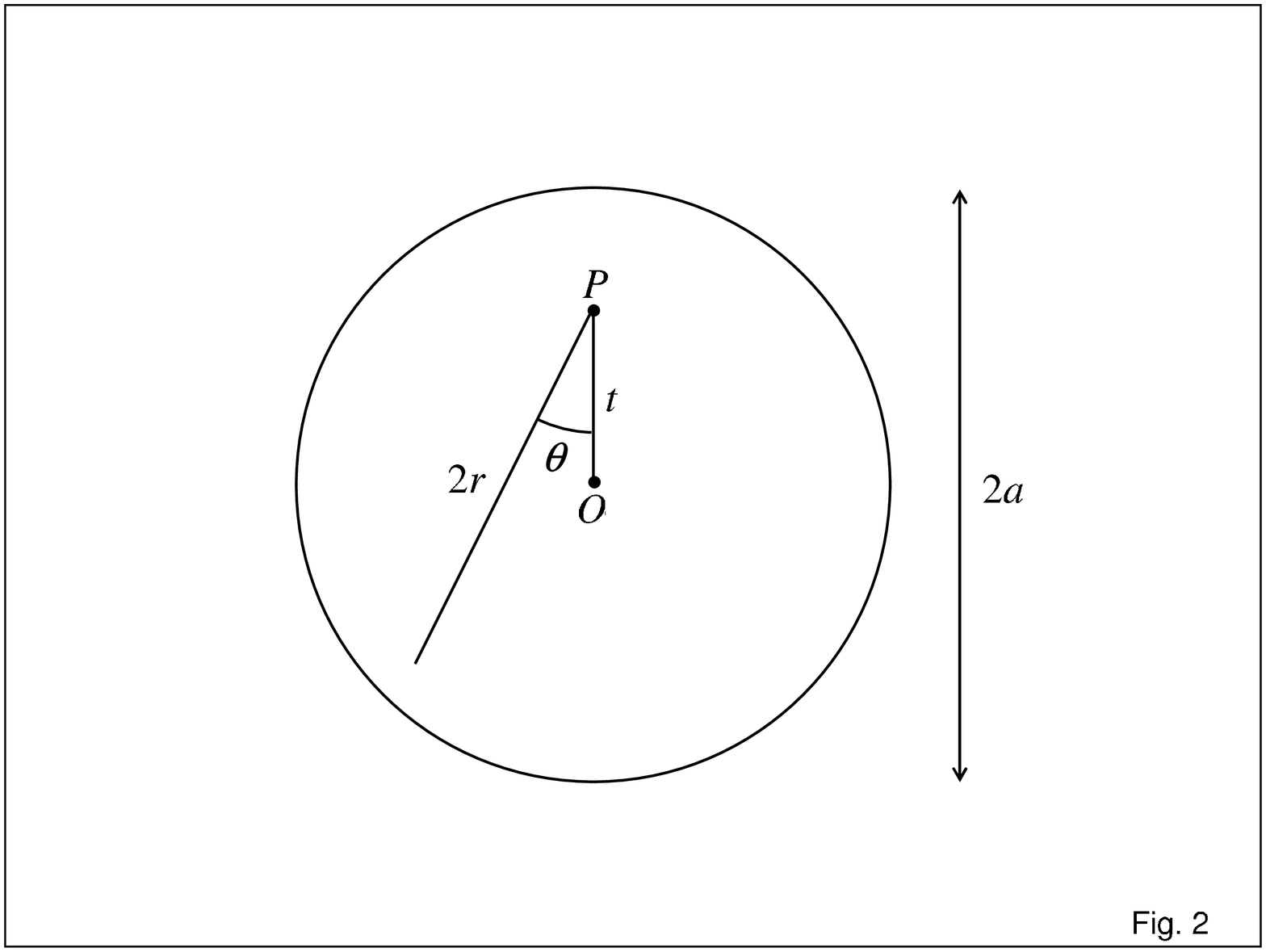, angle=270, width=380pt}
      \caption{Line segment (i.e. chord of the hairpin) enclosed in a circular cross section.}
    \end{center}
\end{figure}

\begin{figure}[htbp]
  \begin{center}
      \epsfig{file=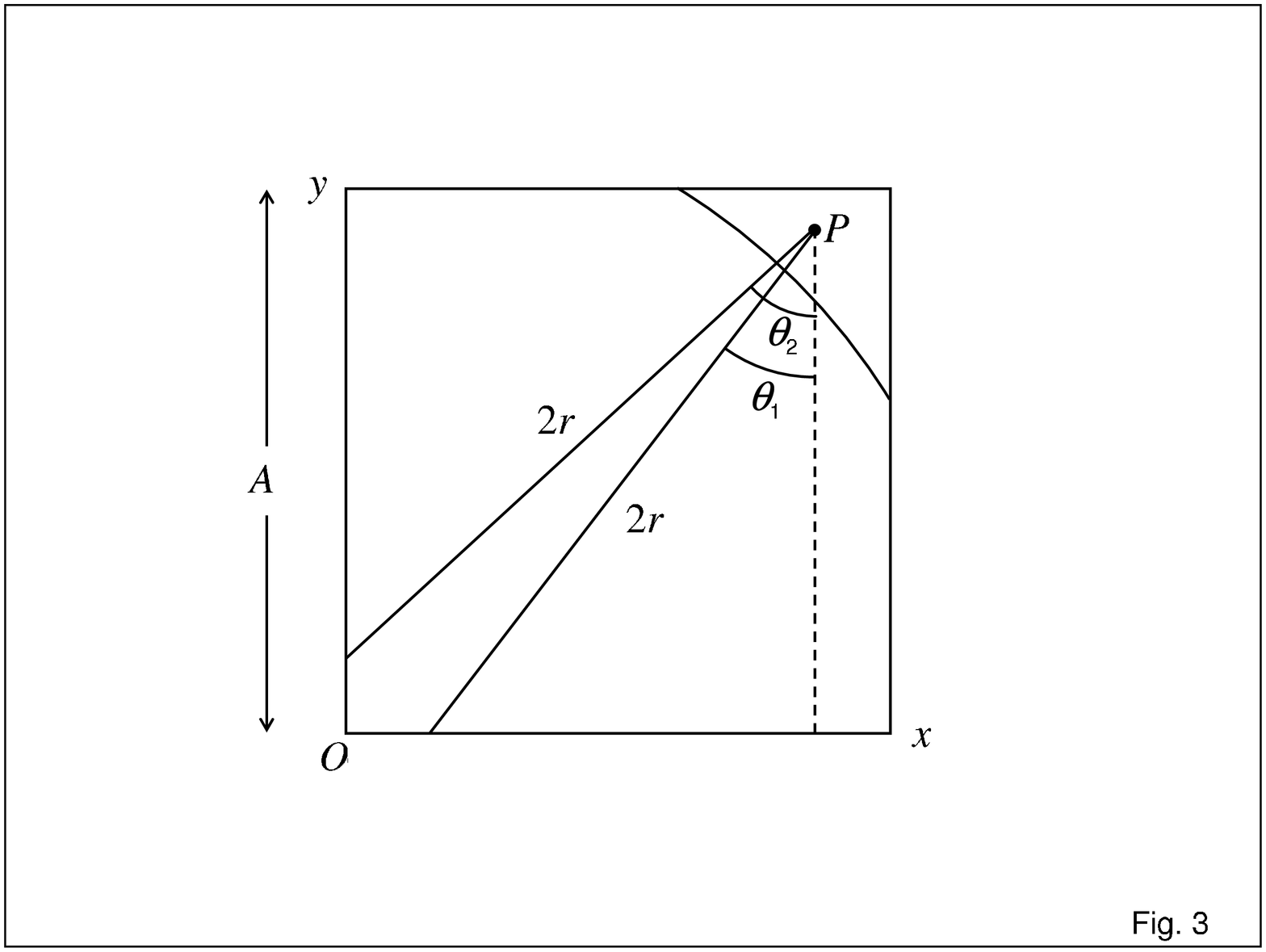, angle=270, width=380pt}
      \caption{Line segment enclosed in a square cross section whose two sides represent the axes of a Cartesian Coordinate system. Point $P$ is restricted to the triangular region.}
    \end{center}
\end{figure}

\begin{figure}[htbp]
  \begin{center}
      \epsfig{file=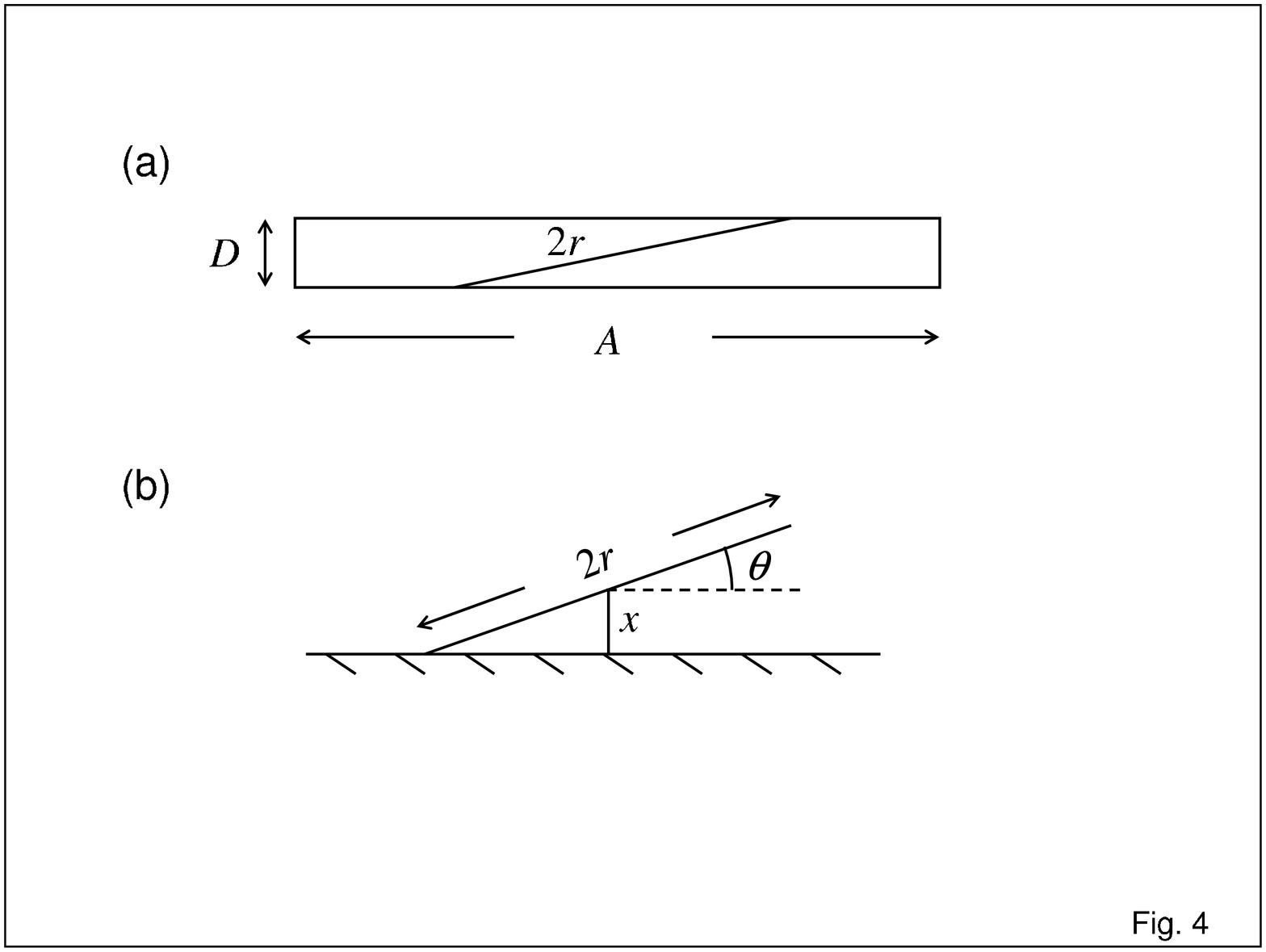, angle=270, width=380pt}
      \caption{ a) Line segment enclosed in the cross section of a narrow slit. b) One end of the segment touches the lower wall whereas its center of mass is fixed at a distance $x$ from the wall.}
    \end{center}
\end{figure}


\begin{thebibliography}{99}
\bibitem{CAS1} E.F. Casassa, J. Polym. Sci. Part B \underline{5}, 773 (1967)

\bibitem{CAS2} E.F. Casassa.and Tagami, Macromolecules \underline{2}, 14
(1969)

\bibitem{COL} C.K. Colton, C.N. Satterfield and C.J. Lai, AIChEJ. \underline{%
21}, 289 (1975)

\bibitem{DAO} M. Daoud and P.G. Gennes, J. Physique \underline{38}, 85 (1977)

\bibitem{CAN} D.S. Cannell and F. Rondelez, Macromolecules \underline{13},
1599 (1980)

\bibitem{GUI} G. Guillot, L. Leger and F. Rondelez, Macromolecules
\underline{18}, 2531 (1985)

\bibitem{ODI1} T. Odijk, Macromolecules \underline{16}, 1340 (1983)

\bibitem{GOR} A.A. Gorbunov and A.M. Skvortsov, Adv. Coll. Int. Sci.62, 31
(1995)

\bibitem{GUS} L.J. Gus, X. Cheng and C.F. Chou, Nanoletters \underline{4},
69 (2004)

\bibitem{TEG1} J.O. Tegenfeldt, C. Prinz, H. Cao, RL. Huang, R.H. Austin,
S.Y. Chou, E.C. Cox and J.C. Sturm, Anal. Bionanal. Chem. \underline{378},
1678 (2004)

\bibitem{BAK} O.B. Bakajn, T.A.J. Duke, C.F. Chou, S.S. Chan, R.H. Austin
and E.C. Cox, Phys. Rev. Lett. \underline{80}, 2737 (1998)

\bibitem{TEG2} J.O. Tegenfeldt, C. Prinz, H. Cao,S. Chou, W.W. Reisner, R.
Riehn, Y.M. Wang, E.C. Cox, J.C. Sturm, P. Silberzan and R.H. Austin, Proc.
Natl. Acad. Sci. \underline{101}, 10979 (2004)

\bibitem{REI} W. Reisner, K.J. Morton, R. Riehn, Y.M. Wang, Z. Yn, M. Rosen,
J.C. Sturm, S.Y. Chou, E. Frey and R. H. Austin, Phys. Rev. Lett. \underline{%
94}, 196101 (2005)

\bibitem{CHO} M.C. Choi, C.D. Santangelo, O, Pelletier, J.H. Kim, S.Y. Kwom,
Z. Wen, Y. Li, P.A. Pincus, C.R. Safinya and M.W. Kim, Macromolecules
\underline{38}, 9882 (2005)

\bibitem{KOS} S. K\"{o}ster, D. Steinhauser and T. Pfohl, J. Phys. Cond.
Mat. \underline{17}, S 4091 (2005)

\bibitem{MAN} J.T. Mansion, C.H. Reccins, J.D. Cross and H.C. Craighead,
Biophys. J. \underline{90}, 4538 (2006)

\bibitem{FAN} R. Fan, R. Karnik, M. Yue, D. Li, A. Majumdar and P. Yong,
Nanoletters, \underline{5}, 1633 (2005)

\bibitem{BAL} A. Balducci, P. Mao, J. Han and P.S. Doyle,Macromolecules,
\underline{39}, 6273 (2006)

\bibitem{RIE} R. Riehn, M. Lu, Y. Wang, S.F. Lim, E.C. Cox and R.H. Austin,
Proc. Natl. Acad. Sci, \underline{102}, 10012 (2005)

\bibitem{RAN} G.C. Randall, K.M. Schultz and P.S. Doyle, Lab Chip.
\underline{6}, 516 (2006)

\bibitem{JO} K. Jo, D.M. Dhingra, T. Odijk, R. Runnheim, D. Forrest and D.C.
Schwartz, "A single molecule barcoding system in nanoslits for genomic
analysis", preprint

\bibitem{KHO} A.R. Khokhlov and A.N. Semenov, J. Phys. A \underline{15},
1361 (1982)

\bibitem{GEN} P.G. de Gennes, in "Polymer Liquid Crystals",eds. A. Ciferri,
W.R. Krigbaum and R.B. Meyer, Academic, N.Y. 1982

\bibitem{WAR} M. Warnar, J.M.F. Gunn and A. Baumg\"{a}rtner, Phys. A
\underline{18}, 3007 (1985)

\bibitem{GLA} M.L. Glasser, V. Privman and A.M. Szpilka, J. Phys A
\underline{19}, L 1185 (1986)

\bibitem{VRO} G.J. Vroege and T. Odijk, Macromolecules, \underline{21}, 2848
(1988)

\bibitem{ODI2} T. Odijk, J. Chem. Phys. \underline{105}, 1270 (1996)

\bibitem{SPA} A.J. Spakowitz and Z.G. Wang, J. Chem. Phys. \underline{119},
13113 (2003)

\bibitem{DIJ} M. Dijkstra, D. Frenkel and H.N.W. Lekkerkerker, Physica A
\underline{193}, 374 (1993)

\bibitem{BUR} T.W. Burkhardt, J. Phys. A \underline{30}, L 167 (1997)

\bibitem{BIC} D.J. Bicout and T.W. Burkhardt, J. Phys. A \underline{34},
5745 (2001)

\bibitem{CHE} J.Z.Y. Chen, D.E. Sullivan and X. Yuan, Europhys. Lett.
\underline{72}, 89 (2005)

\bibitem{YAM} H. Yamakawa and W.H. Stockmayer, J. Chem. Phys. \underline{57}%
, 2843 (1972)

\bibitem{DAN} H.E. Daniels, Proc. Camb. Phil. Soc. \underline{37}, 244 (1941)

\bibitem{RUB} R.J. Rubin, J. Chem. Phys. \underline{56}, 5747 (1972)

\bibitem{WEI} G.H. Weiss and R.J. Rubin, J. Stat. Phys. \underline{14}, 333
(1976)

\bibitem{SHI} J. Shimada and H. Yamakawa, Biopolymers \underline{27}, 657
(1988)

\bibitem{CLA} M.M.A.E. Claessens, M. Bathe, E. Frey and A.R. Bausch, Nature
Materials \underline{5}, 748 (2006)

\bibitem{BUR2} T.W. Burkhardt, J. Phys. A, \underline{28}, L629 (1995)
\end{thebibliography}
\end{document}